\documentclass[a4paper, draftcls, onecolumn]{IEEEtran}
\IEEEoverridecommandlockouts
\usepackage{cite, url}
\usepackage{amsmath,amssymb,amsfonts, multirow, array, hyperref}
\usepackage{algorithmic}
\usepackage{graphicx}
\usepackage{textcomp}
\usepackage{booktabs}
\usepackage{xcolor}
\newcolumntype{C}[1]{>{\raggedright\arraybackslash}m{#1}}
\def\BibTeX{{\rm B\kern-.05em{\sc i\kern-.025em b}\kern-.08em
    T\kern-.1667em\lower.7ex\hbox{E}\kern-.125emX}}

\begin{document}

\title{Six Key Enablers for Machine Type Communication in 6G}

\author{
Nurul Huda Mahmood{\small $^{1}$}\thanks{This work has been submitted to the IEEE for possible publication. Copyright may be transferred without notice, 
after which this version may no longer be accessible},
Hirley Alves{\small $^{1}$},
Onel Alcaraz L\'{o}pez{\small $^{1}$},
Mohammad Shehab{\small $^{1}$}, \\
Diana P. Moya Osorio{\small $^{1, 2}$}, and 
Matti Latva-aho{\small $^{1}$}\\
\fontsize{9}{9}\selectfont\itshape
$~^{1}$
{\href{https://www.oulu.fi/6gflagship/}{6Gflagship.com}}, University of Oulu, Finland.\\
$~^{2}$
Department of Electrical Engineering, Federal University of S\~{a}o Carlos, S\~{a}o Carlos, SP, Brazil.\\
$~^{}$Emails: \{firstname.lastname\}@oulu.fi\\
}

\maketitle
\thispagestyle{empty}

\begin{abstract}

While 5G is being rolled out in different parts of the globe, few research groups around the world $-$ such as the Finnish 6G Flagship program $-$ have already started posing the question: \textit{What will 6G be?} The 6G vision is a data-driven society, enabled by near instant unlimited wireless connectivity. Driven by impetus to provide vertical-specific wireless network solutions, machine type communication encompassing both its mission critical and massive connectivity aspects is foreseen to be an important cornerstone of 6G development. This article presents an over-arching vision for machine type communication in 6G. In this regard, some relevant performance indicators are first anticipated, followed by a presentation of six key enabling technologies. 
\end{abstract}


\section{Introduction}
\label{sec:introduction}
The fifth-generation (5G) cellular standard, also known as 5G New Radio (NR), is designed to serve the growing demand of multi-service wireless communication. 5G NR ushered in a paradigm shift by introducing `Ultra-Reliable and Low Latency Communications' (URLLC) and `massive Machine Type Communications' (mMTC) service classes, thus facilitating the coexistence of services with highly heterogeneous requirements. URLLC addresses critical applications in different vertical industries, whereas mMTC entails providing energy- and spectral-efficient connectivity to a large number of Internet of Things (IoT) devices.

State of the art URLLC and mMTC solutions can be viewed from two different perspectives. The first takes a system design approach by improving existing networks to support the emerging requirements. Examples include physical layer redesign~\cite{3gppTS38300}, scheduling and resource allocation aspects~\cite{karimi_ieeeAccess2018} and IoT-oriented radio technologies like Narrowband IoT. The second perspective, such as finite blocklength transmission~\cite{polyanskiy_trIT2010} and coded random access~\cite{PSL+15_comMag}, are primarily motivated by the information-theoretic aspects of these service classes. 

However, both of this approaches leave a number of loose ends with respect to fully complying with the envisioned requirements of 5G NR. Meanwhile, new requirements continue to emerge with the appearance of new vertical applications opened up by 5G NR. Therefore, it is important to start evaluating 5G NR and predicting the evolution of wireless networks towards $2030$. Focusing on this goal, several research projects around the globe have recently started exploring `beyond 5G'/`sixth generation (6G)' networks~\cite{david_6g_2018, KML18_6G_latincom, BML+18_wirt, strinati_6g_2019}. 

The world's first such initiative is the Finnish 6G Flagship program, a recently formed academic and industrial consortium aiming at developing key enabling technologies for 6G~\cite{KML18_6G_latincom}. The 6G vision is a data-driven society, enabled by near instant unlimited wireless connectivity. In order to achieve this vision, future wireless networks are expected to support a wide range of heterogeneous and sometimes conflicting requirements.

Driven by impetus to provide vertical-specific wireless network solutions, machine type communication (MTC) will be an important cornerstone of 6G development. Many MTC/IoT applications encompass aspects of both, URLLC and mMTC, service classes. Therefore, these service classes are expected to be under the same umbrella in 6G, to be distinguished as critical MTC (cMTC)/mMTC, respectively.  Improving the network scalability, reliability, latency and efficiency in terms of spectral usage and energy consumption while decreasing the deployment costs, are the main design goals for an MTC optimized, cost-effective network.

%Some trends and applications have already emerged looking forward to 6G, for instance, terahertz (THz) data communication networks, self-learning and autonomous networks with the help of artificial intelligence (AI), Tactile Internet, Internet of Skills, holographic communications with enhanced and faster radios than those of 5G NR with even higher throughput, and wireless augmented reality/virtual reality.

\subsection*{Key Performance Indicators for MTC in 6G}
Reliability, latency, device density and energy efficiency are among the main key performance indicators (KPI) pertinent to MTC. The 6G reliability and latency requirement is expected to be diverse and use case specific, with the most extreme values being $10^{-9}$ and $0.1$ millisecond (ms), respectively, corresponding to the current requirements for wired industrial control networks in Industry 4.0 applications~\cite{BML+18_wirt}. 

Extrapolating the ITU specified 5G device density of \textit{a million IoT devices per km$^2$} and considering that 3D connectivity will be an important capacity measure in 6G~\cite{strinati_6g_2019}, we anticipate networks required to support about {$10$ devices per m$^{2} - 100$ devices per m$^{3}$.}

In terms of energy efficiency, 6G will introduce ultra-long battery life aided by a combination of energy-efficient communication, advanced battery technology and energy harvesting (EH) techniques. The ultimate vision is to completely remove the need for separately charging mobile devices~\cite{david_6g_2018}. 

In addition, 6G is expected to be the first generation to consider from scratch a number of new requirements covering technical (e.g., positioning accuracy), societal (e.g., measuring \textit{connectivity as a human right}), environmental (e.g., CO$_2$ footprint) and economical aspects. 

\subsection*{Major Contributions}
\begin{figure}[htb] 
	\centering
	\includegraphics[width=0.7\columnwidth]{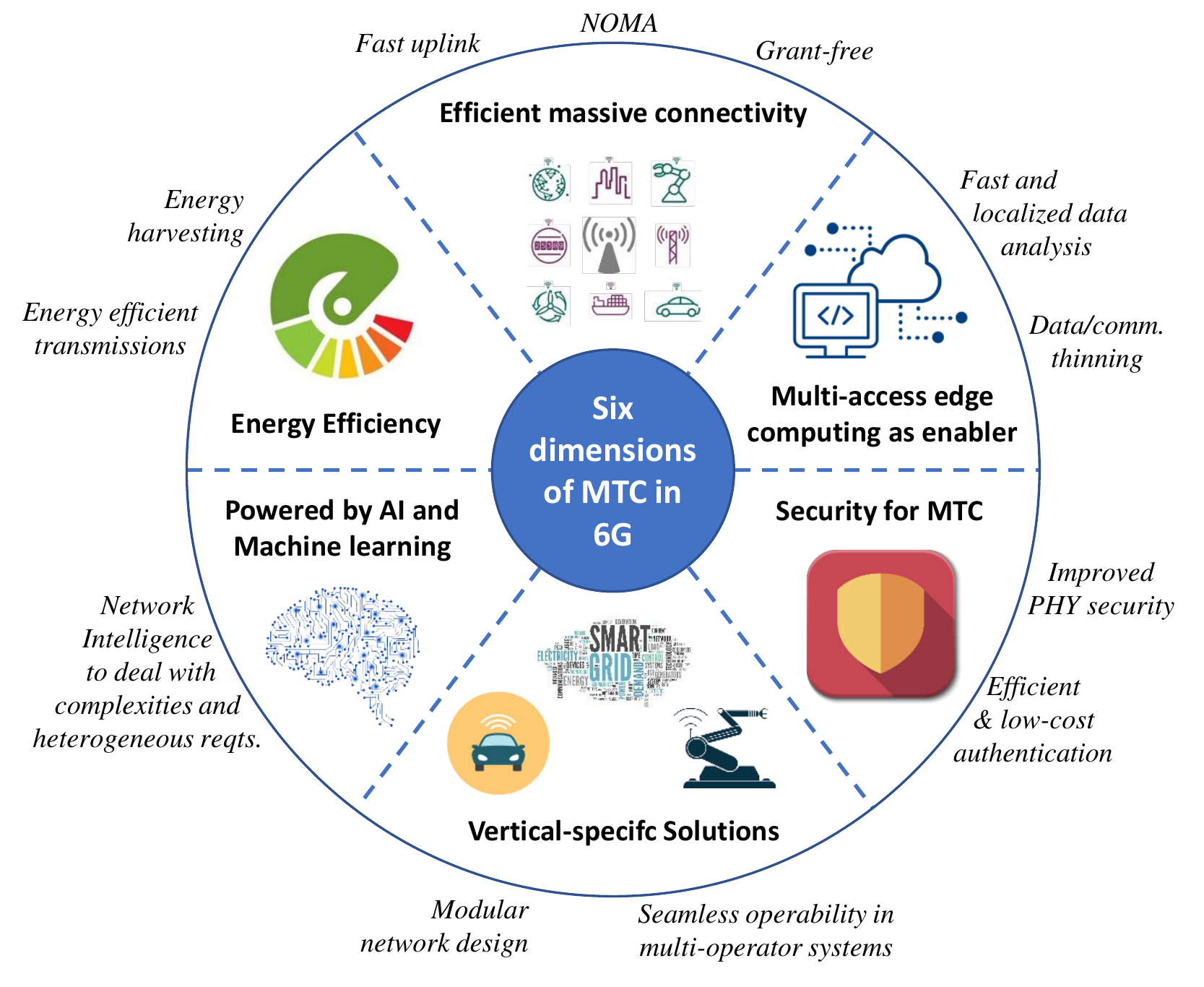}
	\centering
	\caption{Six key enablers for MTC in 6G and their respective solution components.}
	\label{fig:overview}
\end{figure}

This paper presents an over-arching vision for MTC in 6G. In particular, we discuss six key technologies that will enable the ambitious design requirements outlined above. These six enablers and corresponding solution components are illustrated in Figure~\ref{fig:overview}, and detailed in the rest of this article. 

We foresee that some existing elements like massive connectivity and energy efficiency will continue to evolve and be further optimized. Additionally, new security considerations and the applications of emerging technology components like multi-access edge computing (MEC) and machine learning (ML) are highly relevant. Within these aspects, topics like network slicing, edge intelligence and ML enabled algorithms will play prominent roles. 

With the growing need to provide vertical-specific wireless connectivity solutions, different vertical sectors will be intricately involved in 6G design and development. Consequently, ensuring seamless interoperability across different systems within the same vertical will be an important design challenge. 

The six enablers presented are by no means exhaustive. Other potential important enablers for 6G not covered in this article include ultra-high speed links exploiting terahertz and visible light communication, utilizing unmanned aerial vehicles and very low-earth orbit satellites as network infrastructure, and the integration of metamaterials and intelligent structures like software-defined materials and fluid antennae with massive MIMO technology.  

% 1075 words so far

%%%%%%%%%%%%%%%%%%%%%%%%%%%%%%%
%%%%%%%%%%%%%%%%%%%%%%%%%%%%%%%
%%%                         %%%
%%%         SECTION         %%%
%%%                         %%%
%%%%%%%%%%%%%%%%%%%%%%%%%%%%%%%
%%%%%%%%%%%%%%%%%%%%%%%%%%%%%%%

\section{Efficient and Fast Massive Connectivity}
\label{sec:massiveAccess}

Ensuring fast, efficient and reliable channel access for a large number of diverse MTC devices with dynamic and generally low payload traffic and varying latency/reliability requirements is a key design challenge for MTC. The conventional access mechanism of granting exclusive rights to users through a four-way handshake procedure is not suitable for URLLC/mMTC services due to their diverse and challenging requirements. Therefore, efficient radio access technology (RAT) solutions are imperative. 

In the physical layer (PHY), using agile numerology can improve the latency and reliability of wireless links. For instance, 5G NR has introduced mini-slots of duration as low as $0.07$ ms as opposed to the fixed slot duration of one ms in LTE, thereby considerably reducing the minimum transmission time~\cite{3gppTS38300}. Reducing the time it takes to establish channel access is also important for latency reduction. In this respect, grant-free (GF) random access allows users to transmit immediately upon data packet arrival at PHY, thereby reducing the access latency and signaling overhead. 

As URLLC and mMTC services are becoming more interlinked moving towards 6G, novel unified solutions and further optimization of existing technologies are required to meet the challenges of efficient and fast massive connectivity. Three potential research directions are hereby presented.

\subsection*{Predictive Resource Allocation and Scheduling}
The inherent properties of MTC traffic~\cite{ANS19_fastUL} can be utilized to allocate resources pre-emptively. For example, correlation in the traffic arrival among neighboring transmission nodes can be harnessed to allocate resources to a given node conditioned on the neighbor's transmission. Alternately, semi-persistent scheduling~\cite{3gppTS38300} can be efficient for use cases with periodic traffic arrival. 

\begin{figure}[htb] 
	\centering
	\includegraphics[width=0.7\columnwidth]{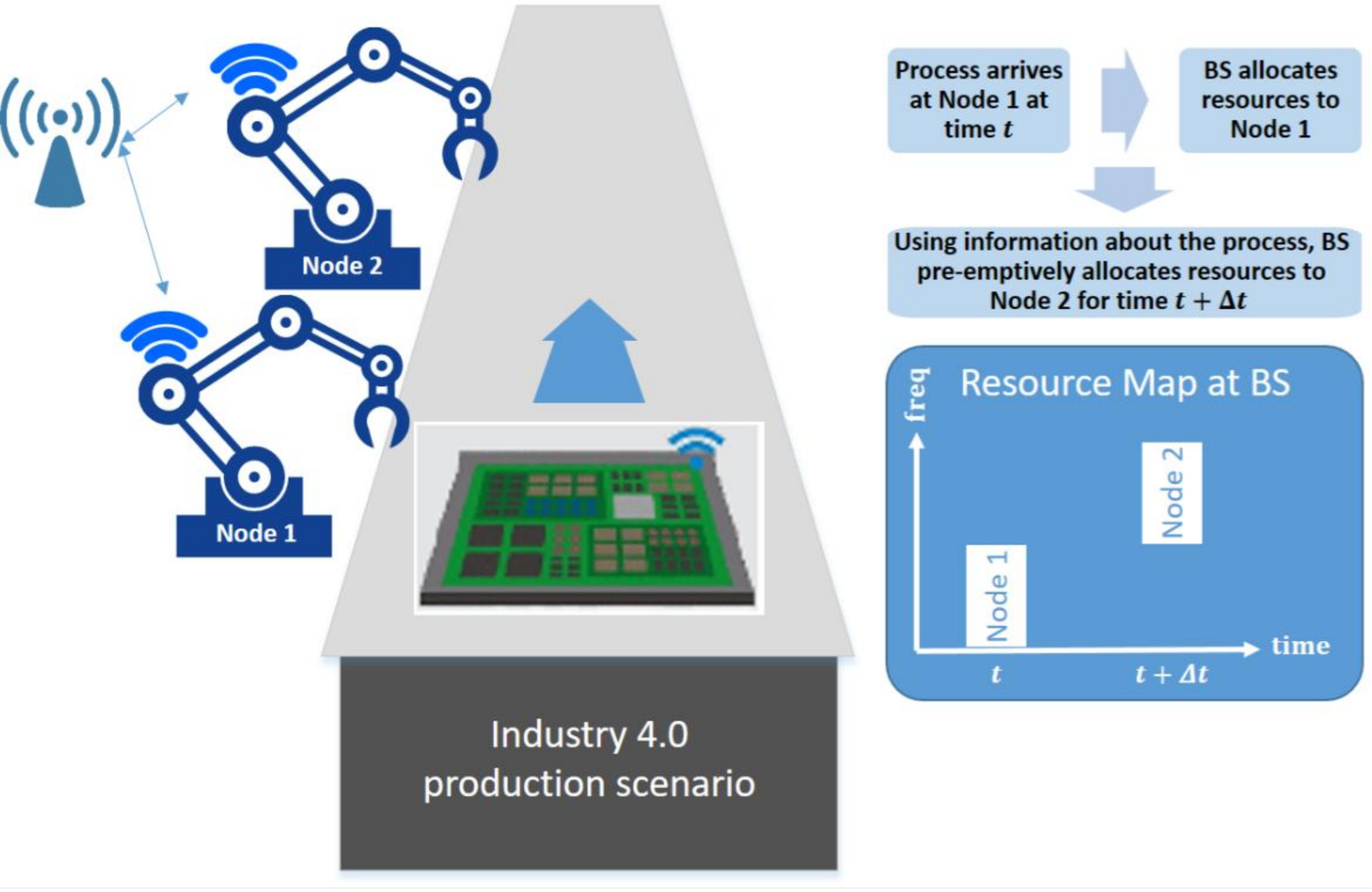}
	\centering
	\caption{An illustration of preemptive resource allocation using information about the traffic correlation between neighboring nodes in an industry 4.0 scenario.}
	\label{fig:preEmptive}
\end{figure}

An illustration of a pre-emptive scheduling is shown in Figure~\ref{fig:preEmptive}, where we consider an industrial process involving two manufacturing nodes operating sequentially. Assuming the time difference for the manufacturing process between the two nodes is $\Delta t$, the base station (BS) can pre-emptively allocate resources to the second node at time $t + \Delta t$ when a process arrives at the first node at the time $t$.

\subsection*{Enhancement of GF Random Access}
Controlling and resolving collisions is one of the main challenges in GF random access. The use of advanced reception techniques like non-orthogonal multiple access (NOMA) and successive interference cancellation in cooperation with GF transmissions enhances its reliability. Moreover, GF can be combined with conventional schemes. For instance, the initial transmission can be performed over dedicated grant-based resources, while proactive hybrid automatic repeat request retransmissions can be performed over shared resources by using GF schemes. In this context, differentiated random access and scheduling policies considering the latency budget will gain increasing attention. 

\subsection*{Improved NOMA Schemes}
NOMA allows transmissions to be multiplexed in time by relying on advanced receivers to detect overlapping transmissions. Most NOMA techniques can be grouped into signature-domain or power-domain multiplexing. Differentiation between users sharing the same resource is done based on the use of different signatures in the former, and by utilizing difference in their received signal power in the latter. NOMA studies in 6G are likely to focus on improving transmitter side processing for the signature domain NOMA and receiver side processing for the power domain NOMA. 

% about 1230 + 590 words so far

%%%%%%%%%%%%%%%%%%%%%%%%%%%%%%%
%%%%%%%%%%%%%%%%%%%%%%%%%%%%%%%
%%%                         %%%
%%%         SECTION         %%%
%%%                         %%%
%%%%%%%%%%%%%%%%%%%%%%%%%%%%%%%
%%%%%%%%%%%%%%%%%%%%%%%%%%%%%%%

\section{Security for MTC}
\label{sec:security}
With a massive number of connected devices, huge bandwidth increase and exploding use cases from various vertical industries, 6G will unavoidably introduce enormous security challenges. This is particularly challenging for MTC considering the device constraints, which are usually delay-sensitive and/or limited in hardware, processing capabilities, storage memory, cost, and energy. Therefore, low-cost and efficient solutions for security and privacy aspects will be highly demanding, as current cryptographic methods are unsuitable for most MTC use cases. 
In the following, we highlight some paramount aspects to be addressed for defining secure 6G networks.

\subsection*{Efficient and Low-Cost Authentication and Authorization}
Simultaneous authentication, authorization and accounting (AAA) processes from millions of connected MTC devices can lead to severe signalling congestion. Short payloads and latency constraints of MTC traffic, and the limited computational capabilities of the devices, demand more scalable solutions for authentication processes. Lightweight and flexible solutions like group-based authentication schemes, anonymous service-oriented authentication strategies to manage a large number of authentication requests~\cite{art:ni2018}, lightweight physical-layer authentication, and the integration of authentication with access protocols~\cite{art:pratas2017} represent promising solutions that are likely to be adopted in 6G. 

User identification will also be an important challenge in 6G. Conventional subscriber identity module (SIM) based solution is simply not scalable and cost-effective for billions of IoT devices, and new AAA mechanisms are therefore necessary. 

\subsection*{Network Slice Security}
\begin{figure}[htb] 
	\centering
	\includegraphics[width=\columnwidth]{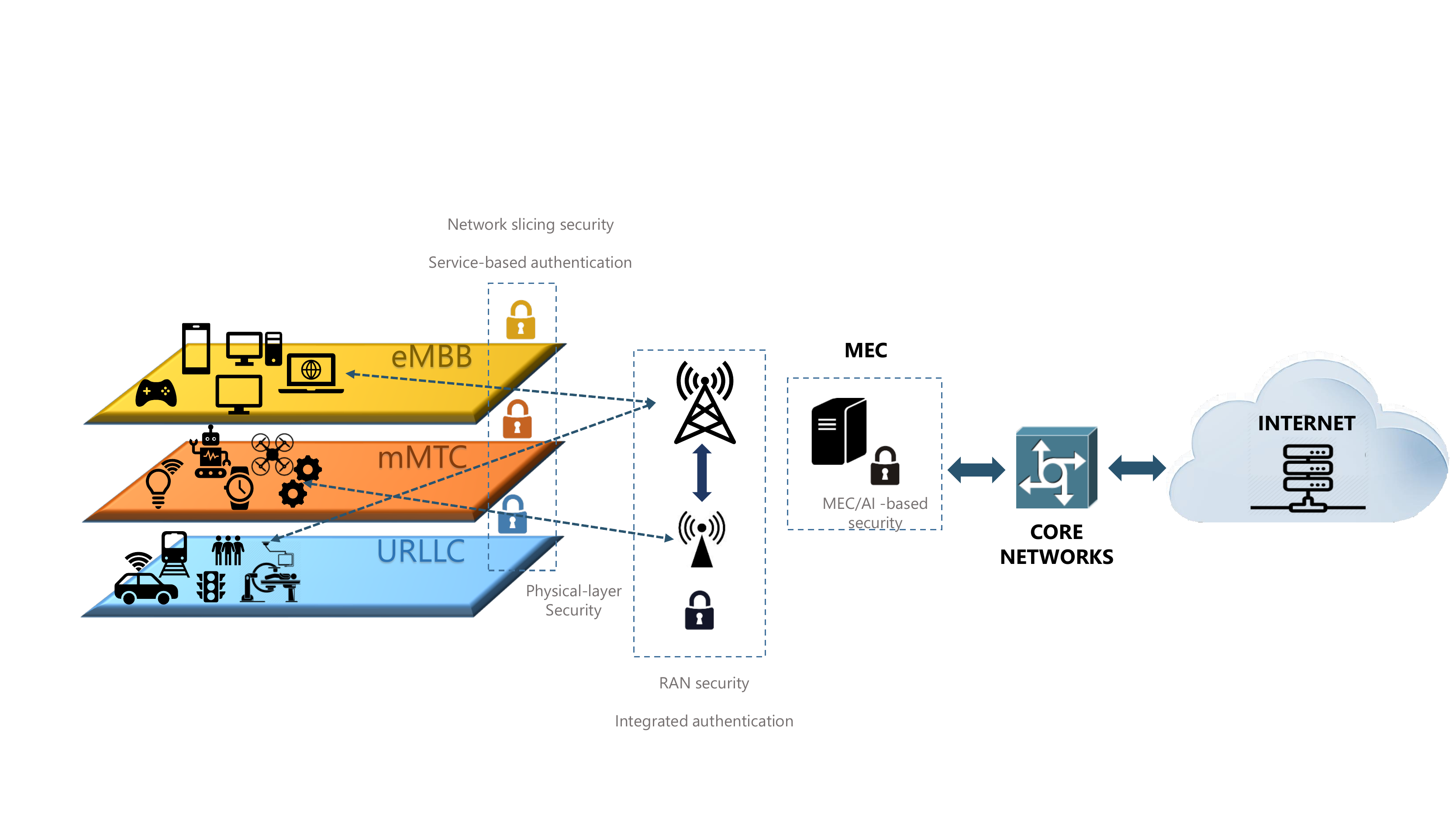}
	\centering
	\caption{Network security solutions for core networks with network slicing architecture.}
	\label{fig:netSlicing}
\end{figure}

Network slicing enables the core network infrastructure to be seamlessly shared across different service classes. The differences in the quality of service (QoS) requirements and security levels of different network slices corresponding to different service classes can be accommodated through discriminated security. An important consideration for network slicing is to ensure secure isolation between the slices such that malicious attacks on a given slice cannot impact the operation of other slices. Moreover, customized security designs with adaptive AAA protocols among different domain infrastructure must be designed~\cite{art:ni2018}. A schematic of the network slicing architecture and different associated security solutions is shown in Figure~\ref{fig:netSlicing}.

\subsection*{Improved Physical Layer Security Techniques}
Physical layer security (PLS) techniques enable secure transmission while preventing eavesdropping by exploiting the wireless channel characteristics. PLS techniques have shown to be a promising alternative for providing security in MTC networks, as they do not rely on complex processes of encryption or decryption. However, it is crucial to develop more robust, efficient and appealing PLS techniques that simultaneously satisfy the requirements on reliability, energy overhead, latency, short payloads, and throughput, and tailored to MTC applications. Moreover, evaluating the performance of secure systems for MTC will demand new secrecy metrics for realistic and practical scenarios compared to the legacy secrecy capacity or secrecy outage probability.

\subsection*{Edge Intelligence to Prevent Security Attacks}
MTC networks are evolving into highly heterogeneous networks, where the wide diversity of devices, capabilities, and services will introduce unprecedented and more powerful forms of artificial intelligence (AI) based malicious security threats. However, computing capabilities provided by emerging network technologies such as MEC and software-defined networks (SDN), which provide powerful computation capabilities at the edge tier, can help to design effective strategies to prevent from distributed denial of service (DDoS) attacks jointly with advanced ML and data mining strategies~\cite{PP19_DDoSDefense}.

% about 1230 + 590 + 520 = 2340 words so far

%%%%%%%%%%%%%%%%%%%%%%%%%%%%%%%
%%%%%%%%%%%%%%%%%%%%%%%%%%%%%%%
%%%                         %%%
%%%         SECTION         %%%
%%%                         %%%
%%%%%%%%%%%%%%%%%%%%%%%%%%%%%%%
%%%%%%%%%%%%%%%%%%%%%%%%%%%%%%%

\section{Powering Massive Deployments Through Wireless Energy Transfer}
\label{sec:energy}
Improved battery lifetime, especially for IoT devices, was a key design focus in 5G NR. Powering a large number of connected devices in an efficient and green way while guaranteeing uninterrupted operation will continue to be an important design challenge in 6G. The ultimate vision is to completely remove the need for separately charging mobile devices, enabled by a combination of energy-efficient communication, advanced battery technology, offloading of energy-intensive processing to the edge (see Section~\ref{sec:MEC}) and EH techniques. EH is an efficient solution to avoid replacing or externally recharging batteries, a procedure that may be costly or impossible in hazardous environments, building structures or inside the human body. 

\subsection*{Energy Harvesting Sources}
\label{sub:EHsources}

The sources of harvestable energy can be from natural sources like solar, vibrational, thermal, biological (such as blood pressure), microbial fuel cells (bio-electrochemical transducers that convert microbial reducing power into electrical energy), human powered (such as walking) etc. Furthermore, energy can be harvested from man-made sources via wireless energy transfer (WET), where energy is transferred from a source to a destination through the transmission of dedicated energy beacon~\cite{UYE+15_EH_JSAC}. %Undoubtedly, EH will be an important enabler of 6G networks by improving the energy efficiency and longevity of devices while enabling future massive deployments. 
In the rest of this section, we present some of the research challenges in the application of EH using WET as an example. 

%\subsection*{Wireless Energy Transfer}
%\label{sub:wet}
%
%The transfer of energy through radio frequency (RF), also referred to as WET, %is an alternate source for powering EH devices. Being wireless 
%allows battery charging operations without physical cable connection. Moreover, it is readily available in the form of transmitted energy, of low cost and form factor, and increases the durability and reliability of the end devices due to its contact-free design~\cite{Niyato.2017}. 

%However, the coexistence of WET with wireless information transfer (WIT) is a key challenge. We next overview some possible solutions addressing this aspect. 
%
\subsection*{Seamless Integration of Wireless Information and Energy Transfer}
\label{sub:integration}

A key challenge in WET is its coexistence with wireless information transfer (WIT). The efficiency of the radio circuitry for WET and WIT are very different. Typical information receivers can operate with very low sensitivities, whereas an EH device needs much more incident power. Thus, energy and information transceivers usually require different RF systems. A potential solution is to have devices with different antenna systems for each application. 

Energy and information transmissions can be performed either in an out-of-band or in-band manner. While the former approach allows avoiding interference, the latter alleviates the spectrum efficiency issue by allowing information and energy to be transmitted over the same band in a time-division or even full-duplex manner. 

\begin{figure}[th] 
	\centering
	\includegraphics[width=0.7\columnwidth]{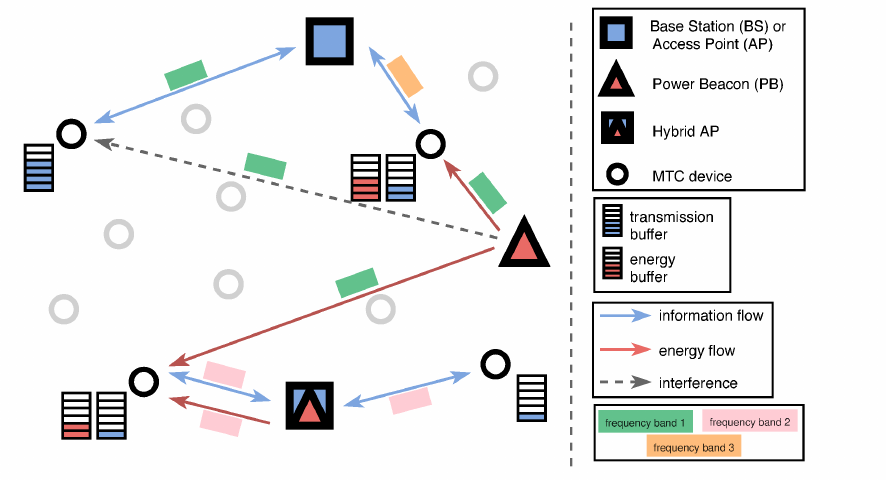}
	\centering
	\caption{Integrating WET into communication architectures. Optimum performance requires taking WET and WIT jointly into account. While one EH MTC device is performing WET and WIT in an out-of-band manner, the other one is exploiting also in-band energy and information transmissions.}
	\label{scenario}
\end{figure}

Incorporating WET into communication architectures as depicted in Figure~\ref{scenario} faces important design challenges. First, the system design must support the coexistence of heterogeneous WIT and WET, e.g., relying on dedicated energy beacons and hybrid WET+WIT access points. Second, WET %aims at powering low-power devices, for which it could be the main or the only energy source, but it 
may also cause some interference to other nearby communicating devices. Therefore, WET design cannot be independent from WIT and a joint optimization of both processes, while considering their particular characteristics, is needed.

In this sense, introducing new metrics, such as effective capacity (EC) and effective energy efficiency (EEE) provides a robust framework for joint WIT and WET optimization. The EC metric can be applied to capture statistical delay requirements in parallel with transmission throughput, whereas EEE is defined as the ratio between EC and the total power consumption. The latter is well-suited to capture the inherent energy-limited characteristics of WET systems and bursty traffic scenarios in MTC.

% about 2340 + 530 = 2870 words so far

%%%%%%%%%%%%%%%%%%%%%%%%%%%%%%%
%%%%%%%%%%%%%%%%%%%%%%%%%%%%%%%
%%%                         %%%
%%%         SECTION         %%%
%%%                         %%%
%%%%%%%%%%%%%%%%%%%%%%%%%%%%%%%
%%%%%%%%%%%%%%%%%%%%%%%%%%%%%%%

\section{Multi-Access Edge Computing}
\label{sec:MEC}

Multi-access Edge Computing is the deployment and operation of distributed computing, caching, network communication, and data analytics resources at the edge of the cellular network. (i.e., locations which are geographically close to the devices that generate and/or use the data). MEC evolved from the fog computing concept, which was mainly introduced so that IoT-like applications could take advantage of cloud computing at the network edge. MEC extends this concept further to encompass additional capabilities like data processing, storage, and using the data to make network decisions~\cite{EFS18_MEClatency}.

\begin{figure}[htb] 
	\centering
	\includegraphics[width=0.7\columnwidth]{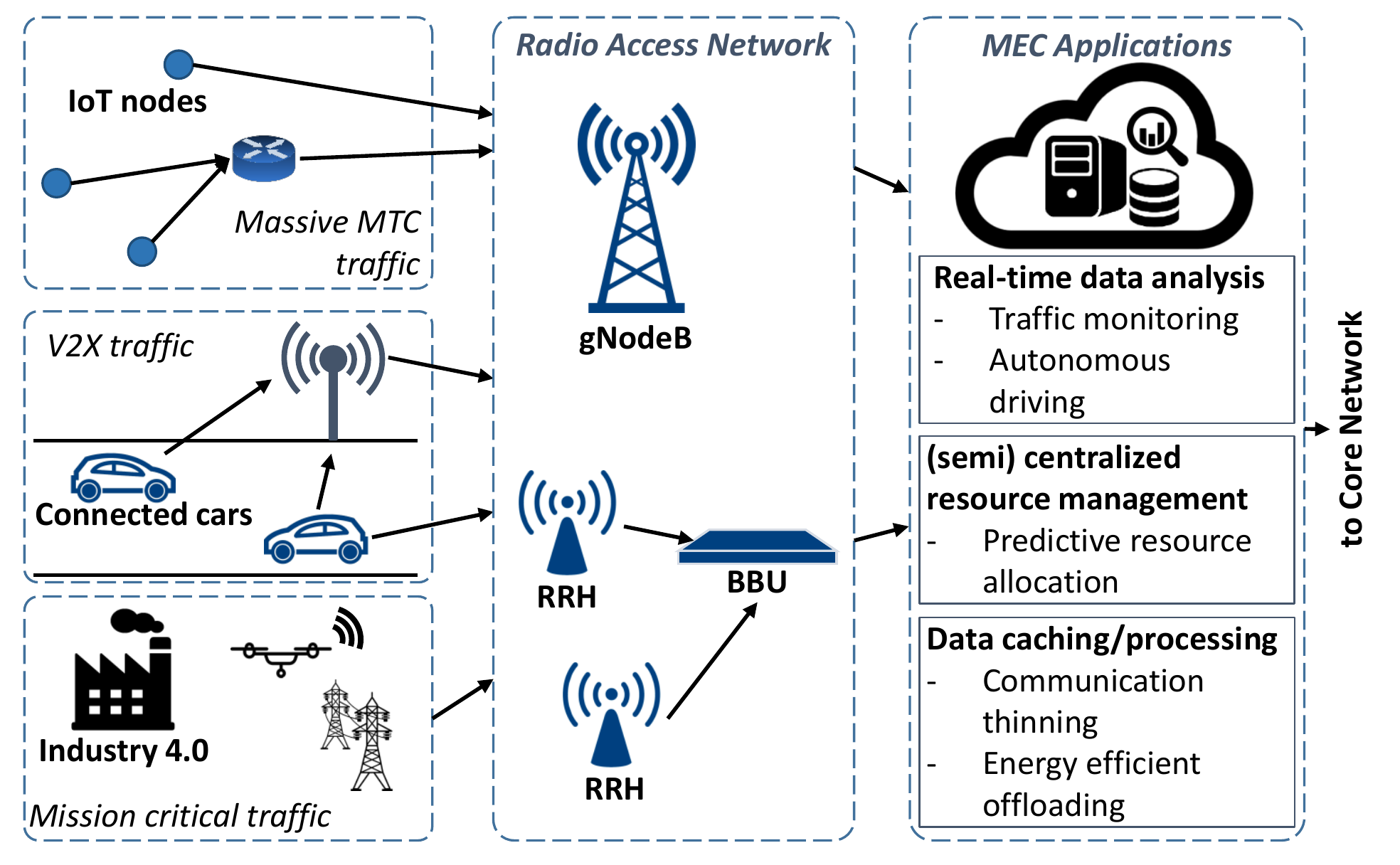}
	\centering
	\caption{Examples of different MEC functionalities as enablers for multi-service communication for massive and critical MTC in 6G.}
	\label{fig:MECfunctions}
\end{figure}

MEC will play a leading role in 6G by operating as an intermediate layer that provides fast and localized data processing for critical and resource constrained applications. Security, vehicle to anything (V2X) communication, energy efficiency and offloading for URLLC can be cited as particularly relevant use cases. Concrete examples of different MEC functionalities as enablers of multi-service communication for massive and critical MTC in 6G are illustrated in Figure~\ref{fig:MECfunctions} and detailed below.

\subsection*{Fast and localized data analysis}
The end-to-end (E2E) latency, which is the latency between the application layers at two ends of a communication link, depends on the access link quality and delays introduced by both the transport and the core network. Many emerging applications like augmented reality and cellular V2X require low E2E latencies. Towards this end, MEC can provide processing capabilities at the edge of the network, thereby significantly reducing the E2E latency~\cite{EFS18_MEClatency}. 

The higher capabilities of the cloud/centralized data centers allow sophisticated and detailed data analysis at the expense of transport and processing delays. The introduction of MEC allows data analytics to be divided into two stages. At the edge, fast and localized data processing, analytics, and content caching can serve critical applications due to its close proximity to end users~\cite{EFS18_MEClatency}, whereas a more in-depth and holistic data analysis over a larger time scale can be carried out at the core. 

The incessant growth of MTC data volume not only increases the amount of data that needs to be processed, but also makes it more difficult to distinguish between useful and unnecessary data. A primary data analysis cycle at the edge, enabled by MEC servers, will allow the huge amount of MTC data generated by end devices to be sorted out and thinned before passing on to the core/cloud network. 

\subsection*{Centralized/semi-centralized resource allocation}
Efficient management of scarce communication and computing resources is a well-known optimization problem. Usually distributed but sub-optimal solutions are implemented in practical networks to limit detailed channel state information (CSI) requirement and complexity. The integration of MEC servers at the network edge in 6G will render (semi) centralized allocation of communication and computing resource on a fast basis practically feasible. Thus, centralized resource allocation algorithms for a cluster of devices connected to the edge can be efficiently executed with limited complexity and CSI. Concurrently, coordination among MEC servers can introduce another layer of resource management over a larger time frame, for example, allocating the number of radio channels to a set of base stations. 

% about 2870 + 530 = 3400 words so far

%%%%%%%%%%%%%%%%%%%%%%%%%%%%%%%
%%%%%%%%%%%%%%%%%%%%%%%%%%%%%%%
%%%                         %%%
%%%         SECTION         %%%
%%%                         %%%
%%%%%%%%%%%%%%%%%%%%%%%%%%%%%%%
%%%%%%%%%%%%%%%%%%%%%%%%%%%%%%%

\section{The Role of Artificial Intelligence and Machine Learning}
\label{sec:AI_ML}

The 6G vision for a data-driven society, enabled by near-instant and unlimited wireless connectivity, will require addressing complex, heterogeneous and often conflicting design requirements. The ever-increasing complexity of the network and its configurations, and the emergence of multi-service communication and applications, demand comprehensive network intelligence. It is imperative to enhance network intelligence to enable self-organization. % (e.g., self-configuration, healing, and optimization). 
Moreover, some applications in 6G may demand dynamic and/or multiple service-type allocation, as opposed to the static categorization into enhanced mobile broadband (eMBB), URLLC and MTC service classes in 5G NR. %Thus, processes such as automatic provisioning on establishing the required network slices for the dynamic service types as well as decentralising the architecture for software defined networking and MEC can be efficiently performed.  

Incorporating network intelligence requires tracking changes in the environment and estimating uncertainties, which are then used as inputs in decision-making and network re-configuration. AI techniques, which aim to equip machines and systems with intelligence, have long been used in specific scenarios to optimize heterogeneous network configurations~\cite{ART:WangACCESS15}. It is once again in the spotlight as a key element in the design of future 6G networks~\cite{strinati_6g_2019}. In this sense, AI tools can solve problems related to observing changes, learning unknowns, identifying issues, forecasting changes, and facilitating the decision-making. 

In this regard, ML is one of the most promising AI tools. ML can be categorized as supervised, unsupervised and reinforcement learning. In supervised learning, the output corresponding to a new input is estimated by learning the input-output relation from exposure to labelled data and their matched outputs. In unsupervised learning, the input data is not labelled \textit{a priori} and the agent needs to find on its own the inherent structure of the input data. Some examples include, clustering of small cell users, and intrusion, fault and anomaly detection in security protocols. Finally, in reinforcement learning, the agent interacts with its environment to achieve its goal without explicitly knowing whether the goal is achieved by attempting to maximize a cumulative reward, such as Markov decision processes and Q-learning in competitive scenarios. 

It is just recently that such tools have started focusing on addressing MTC network issues. Therefore, we expect for the upcoming years that AI-oriented solutions thrive specially in the context of MTC due to its heterogeneous requirements. Moreover, current MTC networks (massive and critical) are static with fixed slicing. With the increase of MEC capabilities and vertical-oriented services, as discussed in Sections~\ref{sec:MEC} and~\ref{sec:verticals}, respectively, we expect MTC network to be autonomous, self-organizing and dynamic. Therefore, they will be able to exploit traffic information and usage patterns of MTC devices and their respective security requirements to perform dynamic slice partitioning and resource allocation according to the needs of each application and/or vertical-oriented service. 

% about 3400 + 470 = 3870 words so far

%%%%%%%%%%%%%%%%%%%%%%%%%%%%%%%
%%%%%%%%%%%%%%%%%%%%%%%%%%%%%%%
%%%                         %%%
%%%         SECTION         %%%
%%%                         %%%
%%%%%%%%%%%%%%%%%%%%%%%%%%%%%%%
%%%%%%%%%%%%%%%%%%%%%%%%%%%%%%%

\section{The Rise of the Verticals}
\label{sec:verticals}

%The Fifth-Generation cellular standard is designed to serve the growing demand of multi-service wireless connectivity. 
In contrast with the primarily user agnostic approach in earlier generations of wireless networks, 5G NR introduced vertical-specific wireless connectivity solutions. The existing approach of having industry-specific standards in different verticals limits scalability, flexibility and cost-effectiveness. While 5G NR has made the first step towards a vertical market driven approach, 6G may be the first wireless standard to completely replace existing industry-specific standards by a single global solution enabling seamless connectivity for the eclectic communication needs across different vertical industry. 

In essence, digitalization via 6G will be strongly driven through the verticals. Connecting various vertical industries through MTC will open up enormous new economical and societal opportunities for the society at large. Alongside, the vertical sectors will be able to capitalize on the communication industry's expertise to enhance their productivity, while the communication industry will have the prospect of new business models in the vertical sectors and be able to find new revenue sources.

\subsection*{Seamless Operability in a Multi-Operator System}
Ensuring compatibility in a multi-operator (including other than legacy telecommunications operators) system will be a major challenge in providing a global vertical-market driven connectivity solution in 6G. The key research question is: how to ensure seamless operability (including, but not limited to, authorization, security, service provisioning, accounting, etc.) across the different information technology (IT) systems of the different players involved? 

Examples of the different players involved in the Automotive sector are the different car manufacturers with their respective V2X solutions (e.g., \textit{Cellular-V2X} advocated by 5G Automotive Association), owners and/or operators of different parts of the road, the agency/agencies managing the spectrum and the national/regional authority/authorities. Persistent health monitoring without compromising security and privacy of a patient as he/she travels across cities or even countries in the eHealth sector is yet another example requiring such seamless inter-operability in a multi-operator system. In this respect, the question of ownership, privacy and security of the generated data is also a significant concern. Who owns the generated data and where should it reside as a device moves across different systems in a given vertical sector? 

Seamless operability in a multi-operator system is not limited to providing connectivity while roaming, rather it concerns enabling compatibility of different IT systems of the different players involved across any given vertical sector. The challenge will be further exacerbated for cMTC application requiring ultra-reliability and low-latency.

\subsection*{Enablers of Vertical-Specific Connectivity Solutions}

Future 6G networks are expected to be flexibly designed to serve the wide-ranging needs of different vertical sectors within a simple and unified system with a modular-designed radio access network connected to a dynamic core network. The modular design of the lower layers will allow addressing the different design requirements of different verticals in an efficient manner. %For instance, separate modules for mMTC, cMTC and eMBB users can be designed. 
On the other hand, the use of network slicing and technologies like network function virtualization (NFV) and SDN will allow the same core network to dynamically serve the needs of different verticals. In this sense, an enhanced RAT, network slicing, MEC, and the use of cloud core are among the key enablers for the traditional wireless communication industry to serve industry-specific communication requirements of different vertical sectors. %The rise of the verticals will therefore be a key driver for 6G development, specifically in the context of mMTC, cMTC and industrial IoT. Digitalization via 6G will therefore be strongly driven through the key verticals.

% about 3870 + 590 = 4460 words so far

%%%%%%%%%%%%%%%%%%%%%%%%%%%%%%%
%%%%%%%%%%%%%%%%%%%%%%%%%%%%%%%
%%%                         %%%
%%%         SECTION         %%%
%%%                         %%%
%%%%%%%%%%%%%%%%%%%%%%%%%%%%%%%
%%%%%%%%%%%%%%%%%%%%%%%%%%%%%%%

\section{Summary and Outlook}
\label{sec:overview}

The introduction of multi-service communication in 5G NR catalyzed a paradigm shift from human-centric communication to MTC. The latter can be broadly categorized into massive MTC $-$ targeting energy and spectrally efficient connectivity solutions for IoT devices, and critical MTC $-$ aimed at applications requiring high reliability and low latency. %With $75$ billion IoT devices expected by $2025$, 
MTC is envisioned to be the primordial focus of the next generation 6G wireless network. In this article, we have forecasted the evolution of MTC in 6G, and presented six technologies that will enable the ambitious design targets. A comprehensive overview of the design challenges, their corresponding enabling technologies and the expected outcomes is outlined in Table~\ref{tab:solution_overview}.

\begin{table*}[t]
\centering
\caption{Overview of the six enabling technologies for MTC in 6G}
\label{tab:solution_overview}
\renewcommand{\arraystretch}{1.3}
\begin{tabular}{p{2cm} p{4cm} p{4cm} p{4cm}}
\toprule
\textbf{Concept} & \textbf{Addressed challenge} & \textbf{Key solution components}& \textbf{Expected outcome}\\
\midrule
%% access
\textit{Efficient massive access} & 
Fast channel access for diverse MTC traffic with low payload &
Fast uplink grant, traffic-aware pre-emptive scheduling, grant-free transmission, NOMA &
Intelligent grant allocation and advanced receivers will allow over $100\%$ channel occupancy
\newline\\
%%Security 
\textit{Security for MTC} &
Ensuring security considering heterogeneous MTC devices, traffic, and service requirements &
Lightweight authentication, authorization and accounting, improved physical layer security, and cross-layer security for network slicing \newline &
Low-cost replacement of SIM-based authentication, enhanced security from sophisticated attacks especially for cMTC services 
\\
%% Energy
\textit{Powering devices through energy harvesting techniques} &
Powering MTC devices in difficult-to-reach locations, enabling ultra-long battery lifetime &
Efficient integration of wireless energy and information transfer, increase of E2E efficiency and the use of effective energy efficiency framework \newline & 
Longer lifetime for MTC devices (even without servicing and maintenance), form factor reduction, green operation 
\\
%% MEC
\textit{Multi-access edge computing enabled solutions} &
MEC as an enabler for massive and critical MTC applications &
Intermediate data analysis at the edge, distributed computing and caching and efficient semi-centralized resource management for resource constrained MTC devices \newline &
Better utilization of scarce resources especially at the lower layers, improved localized performance and data/communication thinning 
\\
%% AI
\textit{The role of artificial intelligence/machine learning} &
Addressing the ever-increasing complexity of the network and its configurations, and the emergence of dynamic multi-service communication &
Intelligent algorithms powered by AI/ML targeting applications that cannot be efficiently solved otherwise due to complexity and other issues \newline &
Improved network performance considering all KPIs 
\\
%% verticals
\textit{Vertical driven network} &
Ensuring seamless operability in a multi-operator system, differentiated services and products for different verticals &
Modular network design at lower layers and the use of networks slicing, NFV, SDN in the higher layers &
Newer markets and revenue sources, improved digitization and productivity of the society\\
\bottomrule
\end{tabular} 

\end{table*}

% about 4460 + 400 = 4860 + 440 (ack + ref) =  words so far

\section*{Acknowledgements}
This work has been performed under the Academy of Finland 6Genesis Flagship program (grant no. 318927). The authors would like to acknowledge the contributions of their colleagues in the project, although the views expressed in this work are those of the authors and do not necessarily represent the project. The work of Diana P. Moya Osorio is partially supported by S\~{a}o Paulo Research Foundation (grant no. FAPESP 2017/20990-6).

%\bibliographystyle{IEEEtran}
%\bibliography{bibliography}

%%%%%%%%%%%%%%%%%%%%%%%%%%%%%%
%==============================
% Bibliography

% Generated by IEEEtran.bst, version: 1.13 (2008/09/30)

%================================
%%%%%%%%%%%%%%%%%%%%%%%%%%%%%%%%%

\begin{IEEEbiographynophoto}{Nurul Huda Mahmood} was born in Chittagong, Bangladesh. He is a Senior Research Fellow within the \href{https://www.oulu.fi/6gflagship/}{\em 6G Flagship} program and is affiliated with Centre for Wireless Communications, University of Oulu, Finland (CWC). His research interests include resource optimization techniques with focus on URLLC/MTC, and modeling and performance analysis of wireless communication systems.
\end{IEEEbiographynophoto}

\begin{IEEEbiographynophoto}{Assistant Professor Hirley Alves} is the head of the Machine-type Wireless Communications Group and leads the URLLC activities under the \href{https://www.oulu.fi/6gflagship/}{\em 6G Flagship} program at CWC. Assist. Prof. Alves is General Co-Chair of the \href{http://www.6gsummit.com/}{\em $1^{st}$ 6G Summit} and General Chair of \href{http://iswcs2019.org/}{ISWCS'2019}. His group is actively working on massive connectivity and URLLC for future wireless networks. 
\end{IEEEbiographynophoto}

\begin{IEEEbiographynophoto}{Onel L.A. L\'{o}pez} (1989) received the B.Sc. (1st class honors) and MSc. degrees in Electrical Engineering from Central University of Las Villas, Cuba (2013) and from Federal University of Paran\'{a}, Brazil (2017). He is currently conducting PhD studies on future wireless networks at CWC.
\end{IEEEbiographynophoto}

\begin{IEEEbiographynophoto}{Mohammad Shehab} (1989) is resuming his doctoral studies with CWC with focus on Ultra Reliable and Machine Type Communication. He has obtained his B. Sc from Alexandria University (2011), two M. Sc degrees from the Arab Academy (2014) and University of Oulu (2017).
\end{IEEEbiographynophoto}

\begin{IEEEbiographynophoto}{Diana P. M. Osorio} [M'16] is currently an Assistant Professor with the Department of Electrical Engineering, Center for Exact Sciences and Technology, Federal University of S\~{a}o Carlos (UFSCar), S\~{a}o Carlos, SP, Brazil. Her research interests include wireless communications in general, cooperative relaying networks, physical layer security, machine-type communications, and UAV-aided communications.
\end{IEEEbiographynophoto}

\begin{IEEEbiographynophoto}{Matti Latva-aho} received the M.Sc., Lic.Tech. and Dr. Tech (Hons.) degrees in Electrical Engineering from the University of Oulu, Finland in 1992, 1996 and 1998, respectively. From 1992 to 1993, he was a Research Engineer at Nokia Mobile Phones, Oulu, Finland after which he joined CWC. Prof. Latva-aho was Director of CWC during the years 1998-2006 and Head of Department for Communication Engineering until August 2014. Currently he serves as Academy of Finland Professor in 2017 – 2022 and is Director for \href{https://www.oulu.fi/6gflagship/}{\em 6Genesis} - Finnish Wireless Flagship for 2018 - 2026. His research interests are related to mobile broadband communication systems and currently his group focuses on 5G and beyond systems research. Prof. Latva-aho has published 350+ conference or journal papers in the field of wireless communications. He received Nokia Foundation Award in 2015 for his achievements in mobile communications research. 
\end{IEEEbiographynophoto}

\end{document}